\documentclass[twocolumn,floatfix,eqsecnum,nofootinbib]{revtex4-2}
\usepackage{graphicx}
\usepackage{amsmath}
\usepackage{amssymb}
\usepackage{bm}
\usepackage{hyperref}

\usepackage{color}

\begin{document}

\author{C. W. J. Beenakker}
\title{Tangent fermions can restore vacuum stability of discrete-time Dirac models}

\affiliation{Instituut-Lorentz, Universiteit Leiden, P.O. Box 9506, 2300 RA Leiden, The Netherlands}

\date{September 2026}

\begin{abstract}
The Dirac quantum walk (a $1+1$-dimensional space-time discretization of the Dirac equation) has a $2\mu$ mass gap both at the center and at the corner of the quasi-energy--momentum Brillouin zone. Gupta and Short recently noted [Quantum \textbf{9}, 1845 (2025)] that the Dirac vacuum can create a particle-hole pair at the zone corner with the \textit{release} of an energy $2\mu$ to the environment. They removed this vacuum instability at the expense of fermion doubling, the appearance of a second low-energy Dirac cone. Here we show that an alternative discretization scheme, with a tangent rather than a sine dispersion relation, offers stability while retaining a single Dirac cone. The key step is the Cayley transformation from the unit circle of Floquet eigenvalues $e^{-i\varepsilon}$ to the real line of unbounded energies $E=2\tan(\varepsilon/2)$. We compute the Schwinger effect (particle-hole pair creation in a uniform electric field) for tangent fermions and show that the pair-production rate agrees with the continuum result for a single Dirac cone. The zone corner is exactly decoupled if the scalar potential is coupled through the Hermitian generator of the quantum map. If it is coupled as a split operator, in order to preserve exact gauge invariance on the lattice, the corner does contribute --- with a weight that vanishes quadratically with the lattice constants, in contrast to the Dirac quantum walk where the zone-corner instability survives the continuum limit.
\end{abstract}
\maketitle

\section{Introduction}
\label{sec:intro}

Quantum evolution in discrete time steps is shared by quantum walks \cite{Ven12}, by digital quantum simulation based on split-operator (Trotterized) circuits \cite{Fau24}, and by periodically driven condensed matter systems \cite{Mor23}. In all of these the dynamics is generated stroboscopically, by a unitary map ${\cal U}$ (the Floquet operator) applied once per time step $\tau$ to a quantum state. The energy becomes modular: The eigenvalues $e^{-i\varepsilon\tau}$ of ${\cal U}$ define the quasi-energy $\varepsilon$ only modulo the bandwidth $2\pi/\tau$ of the time lattice. (We set $\hbar$ to unity.)

The shift of $\varepsilon$ by $2\pi/\tau$ is a temporal Umklapp process, where the system exchanges energy with the effective Floquet drive. This may lead to an instability, referred to as \textit{Trotter heating} in the context of quantum simulations of many-body systems \cite{Laz14,Ale14,Pon15,Buk15,Hey19,Sie19}. The instability is delayed by energy scale separation: When single-particle energies are much smaller than the bandwidth, energy exchanges of order $1/\tau$ are a rare collective process \cite{Aba16,Mor16,Aba17}.

\begin{figure}[tb]
\centerline{\includegraphics[width=0.9\linewidth]{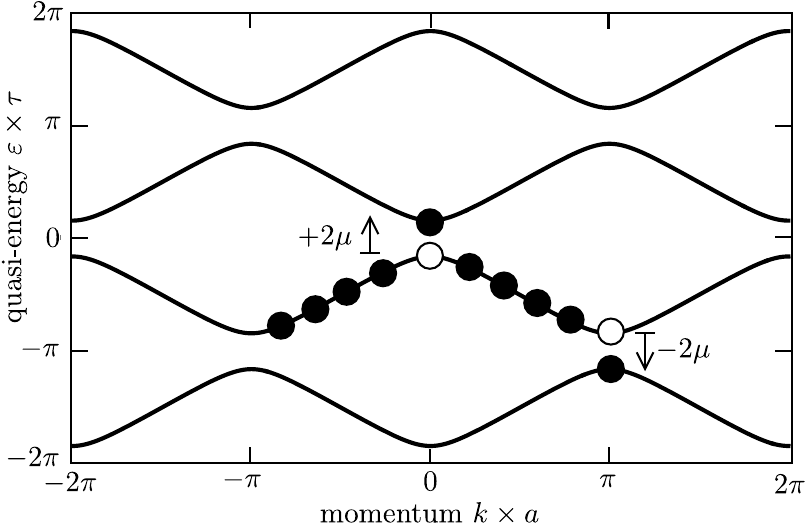}}
\caption{Quasi-energy--momentum dispersion relation \eqref{dispersionDirac} of the Dirac quantum walk \eqref{UDirac}, plotted for mass $\mu=0.5/\tau$ over several Brillouin zones. Black and white dots indicate filled and empty states, respectively. A particle-hole pair created at $k=0$ consumes an energy $2\mu$, while at $k=\pi/a$ it \textit{releases} an energy $2\mu$. The energy balance can be understood as a temporal Umklapp process. While the periodic drive injects an energy $2\pi/\tau$ in the system, the particle at $k=\pi/a$ only gains $2\pi/\tau-2\mu$.
}
\label{fig_extended1}
\end{figure}

The Fermi sea of a Dirac quantum walk does not benefit from this scale separation (see Fig.\ \ref{fig_extended1}). In Feynman's $1+1$-dimensional space-time discretization of the Dirac equation \cite{Fey65}, the mass gap $2\mu$ at $\varepsilon=0$ has a replica at $|\varepsilon|=\pi/\tau$, and particle-hole pair creation processes can occur at both quasi-energies. As pointed out recently by Gupta and Short \cite{Gup25}, pair creation at $|\varepsilon|=\pi/\tau$ \textit{releases} $2\mu$ to the environment via Floquet Umklapp (the periodic drive supplies an energy $2\pi/\tau$ but the system only gains $2\pi/\tau-2\mu$). The time-discretized Dirac vacuum thus becomes unstable to energetically favourable particle-hole pair creation. While the instability appears on the space-time lattice, it does not go away in the continuum limit.

The remedy proposed in Ref.\ \onlinecite{Gup25} is to reduce the energy scale of the Fermi sea, limiting filled states to a quasi-energy range well below $\pi/\tau$. The restored stability comes at the price of fermion doubling: The modified quantum walk \cite{Jol23,Gup26} acquires a second Dirac cone at $\varepsilon=0$. This trade-off is enforced by time reversal symmetry \cite{Bes21,Bee26}: A massless Dirac fermion on a $d+1$-dimensional space-time lattice must have Floquet operator ${\cal U}=\pm 1$ at $2^d$ points in the Brillouin zone (time-reversally invariant momenta), so the removal of a band crossing at $\varepsilon=\pi/\tau$ must be compensated by the appearance of a band crossing at $\varepsilon=0$. In discrete time, it seems, one may have a stable Dirac vacuum or a single Dirac cone, but not both.

Here we present an alternative stabilization, that avoids fermion doubling by replacing the  Floquet update $\Psi(t+\tau)={\cal U}\Psi(t)$ by the Crank-Nicolson update
\begin{equation}
(\openone+\tfrac{1}{2}i\tau{\cal W})\Psi(t+\tau)=(\openone-\tfrac{1}{2}i\tau{\cal W})\Psi(t).\label{CNmap}
\end{equation}
The spectrum of the Hermitian operator ${\cal W}$ is unbounded, its eigenvalues $E\in(-\infty,\infty)$ are related to the quasi-energies $\varepsilon\in(-\pi/\tau,\pi/\tau)$ by the Cayley transformation
\begin{equation}
e^{i\varepsilon\tau}=\frac{1+i E \tau/2}{1-i E \tau/2}\Leftrightarrow E \tau =2\tan(\varepsilon\tau/2),
\label{Evarepsilonrelation}
\end{equation}
a mapping between the real line and the unit circle. We will refer to $E$ as the ``Cayley energy'', to distinguish it from the quasi-energy $\varepsilon$. For ${\cal W}=(2v/a)\sigma_x\tan(ka/2)+\mu\sigma_z$ (spatial lattice constant $a$, velocity $v$) the corresponding dispersion relation has the tangent fermion form \cite{Bee23}
\begin{equation}
E=\pm\sqrt{(2v/a)^2\tan^2 (ka/2)+\mu^2}.
\end{equation}

Since an eigenstate of ${\cal W}$ is an eigenstate of the quantum map \eqref{CNmap}, the Cayley energy is a constant of the motion. The spectrum of $E$ is unbounded, the filled Fermi sea $E<0$ is its ground state, stable to perturbations of ${\cal W}$. The Floquet Umklapp instability is shut off not by scale separation but by the Cayley energy conservation law. The $|\varepsilon|=\pi/\tau$ band crossing survives, as it must to avoid fermion doubling --- but it no longer contributes to the dynamics: The corresponding Cayley energies at $E=\pm\infty$ cannot be coupled to the rest of the spectrum by any bounded perturbation of ${\cal W}$.

This decoupling is exact as a statement about perturbations that enter the generator ${\cal W}$. A scalar potential coupled that way is gauge invariant only up to corrections of order $\tau$, and one may prefer the exactly gauge invariant \textit{split-operator} coupling, in which the potential multiplies the wave function by a phase $e^{-iV\tau/2}$ before and after each step. The two couplings agree in the continuum limit but not on the lattice, and we therefore organize our findings in three levels of decreasing strength:
\begin{enumerate}
\item[(i)] For any bounded perturbation of ${\cal W}$ the Brillouin zone corner is ungapped and exactly decoupled: It sits at infinite Cayley energy, no gap can open there and no transition can reach it. The price is that gauge invariance is recovered only for $\tau\rightarrow0$.
\item[(ii)] For the split-operator coupling in a uniform electric field the corner remains ungapped while supporting an \textit{elastic} channel driven by spectral flow, with scattering probability $\propto(a\tau)^2$; no energy is released to the environment, the vacuum remains stable.
\item[(iii)] A non-uniform potential $\delta V$ in the split-operator coupling opens a gap $\tfrac{1}{2}\mu\tau^2\langle\delta V^2\rangle$ at the Brillouin zone corner, if it varies sufficiently rapidly. This re-opens the Gupta-Short vacuum instability, but now as a lattice effect that vanishes quadratically in the continuum limit $\tau\rightarrow 0$.
\end{enumerate}

The outline of the paper is as follows. In the next section we formulate the tangent fermion quantum map, in the form of a \textit{local} Crank-Nicolson equation \cite{Sta82,Ben83,Pac21,Don22}. The stability for pair creation is studied via the Schwinger effect \cite{Sch51,Gel16,Tay26}, the non-perturbative tunneling across the mass gap induced by an electric field. We compare two ways to couple the scalar potential to the quantum map: In the ${\cal W}$ operator (Sec.\ \ref{sec_Schwinger}, without any zone-corner excitations, but only gauge invariant for $\tau\rightarrow 0$) or as a split operator (Sec.\ \ref{sec:gauge}, exactly gauge invariant but with zone-corner excitations). We conclude in Sec.\ \ref{sec_conclude}.

\section{Two space-time discretizations of the Dirac equation}
\label{sec_instability}

\subsection{Dirac quantum walk}
\label{sec_Dirac}

The Dirac quantum walk on a $1+1$-dimensional space-time lattice has Floquet operator \cite{Fey65}
\begin{equation}
{\cal U}_{\rm Dirac}=e^{-i\mu\sigma_z}e^{-ik\sigma_x},\label{UDirac}
\end{equation}
setting the spatio-temporal lattice constants $a,\tau$ to unity. (We will restore these constants in final equations.) The spinor wave function $\Psi(t)$ evolves in one time step as $\Psi(t+1)={\cal U}_{\rm Dirac}\Psi(t)$. In the continuum limit this reduces to the Dirac equation \cite{Str06,Sun12,Arr14},
\begin{equation}
i\frac{\partial}{\partial t}\Psi(x,t)=(vk\sigma_x+\mu\sigma_z)\Psi(x,t),
\end{equation}
for a spin-1/2 particle of mass $\mu$, momentum $k=-i\partial/\partial x$, and unit velocity $v=1$.\footnote{The Dirac quantum walk has $v=1$ built in, meaning $v=a/\tau$ in terms of the discretization constants. In the tangent fermion quantum map the velocity is a free parameter.} The $\sigma_\alpha$ are Pauli spin matrices.

The eigenvalues $e^{-i\varepsilon}$ of the unitary Floquet operator \eqref{UDirac} define the quasi-energies $\varepsilon$, modulo $2\pi$, dependent on the momentum $k$, modulo $2\pi$. The dispersion relation is given by
\begin{equation}
\cos\varepsilon=\cos k\cos\mu,\label{dispersionDirac}
\end{equation}
plotted in Fig.\ \ref{fig_extended1}.

The quasi-energy band in the range $-\pi<\varepsilon<0$ is identified with the Fermi sea, filled with particles, and the band $0<\varepsilon<\pi$ then represents the empty Dirac vacuum. The bands are separated by a gap $2\mu$ at two points in the first Brillouin zone, at $k=0$ and at $k=\pi$. A particle may tunnel across the gap into the vacuum, leaving behind an empty state (hole) in the Fermi sea. This inelastic process takes an energy $2\mu$ from the environment if it happens at $k=0$, but it \textit{releases} an energy $2\mu$ if it happens at $k=\pi$. 

This is the vacuum instability in discrete space-time of Gupta and Short \cite{Gup25}: Upon coupling to an energy sink, the Dirac quantum walk becomes unstable to the production of energetically favourable particle-hole pairs of high momenta.

\subsection{Tangent fermion quantum map}
\label{sec_tangent}

To avoid the instability we consider the quantum map \cite{Bee23}
\begin{equation}
\begin{split}
&\left(\Phi^\dagger \Phi +\tfrac{1}{2}i{\cal H}\right)\Psi(t+1)=\left(\Phi^\dagger \Phi-\tfrac{1}{2}i{\cal H}\right)\Psi(t),\\
&{\cal H}=v\sigma_x\sin k+\Phi^\dagger\mu\sigma_z\Phi,\;\;
\Phi=\tfrac{1}{2}(1+e^{ik}).
\end{split}\label{CNeq}
\end{equation}
The operator $e^{ik}$ translates a state by one lattice constant, so $\Phi$ averages over adjacent sites.\footnote{A uniform mass $\mu$ commutes with the averaging operator $\Phi$, so we might replace $\Phi^\dagger\mu\Phi\mapsto \mu\cos^2(k/2)$. This expression already shows that the mass becomes ineffective at opening a gap at $k=\pi$.}

Eq.\ \eqref{CNeq} is the space-time discretized Dirac equation of Bender, Milton, and Sharp \cite{Ben83}, reducing to Stacey's generalized eigenproblem \cite{Sta82,Pac21} if only space is discretized. It is a \textit{local} update, of the Crank-Nicolson type $A\Psi(t+1)=B\Psi(t)$ \cite{Cra47}, with sparse matrices $A,B$ (each site is only coupled to its nearest neighbours). The update is ``implicit'', with operators on both sides of the equation.

We may exclude $k=\pi$ from the spectrum by taking an odd number of lattice sites along $x$ with periodic boundary conditions. Then $\Phi$ is invertible and the transformation $\Psi=\Phi^{-1}\tilde{\Psi}$ produces the map $\tilde{\Psi}(t+1)={\cal O}\tilde{\Psi}(t)$ with Floquet operator
\begin{equation}
\begin{split}
&{\cal O}=(\openone+\tfrac{1}{2}i{\cal W})^{-1}(\openone-\tfrac{1}{2}i{\cal W}),\\
&{\cal W}=2v\sigma_x\tan(k/2)+\mu\sigma_z.
\end{split}\label{Omegadef}
\end{equation}
The unitary ${\cal O}$ is the Cayley transform of the Hermitian operator ${\cal W}$. Because ${\cal W}$ is nonlocal (the tangent couples distant lattice sites), the quantum map \eqref{CNeq} implements the local ${\cal H}$ rather than ${\cal W}$.

A comment on the locality of the quantum map: Efficient methods to solve implicit equations of the form $A\Psi(t+1)=B\Psi(t)$, with sparse $A,B$, do not invert $B$ to convert this to an explicit form, since sparsity of the matrices would be lost. Instead, an LU decomposition of $A,B$ allows for a solution of the implicit update at a cost of $N\ln N$ per time step on an $N$-site lattice \cite{Don22}, the same scaling as for explicit split-operator updates.

\begin{figure}[tb]
\centerline{\includegraphics[width=0.9\linewidth]{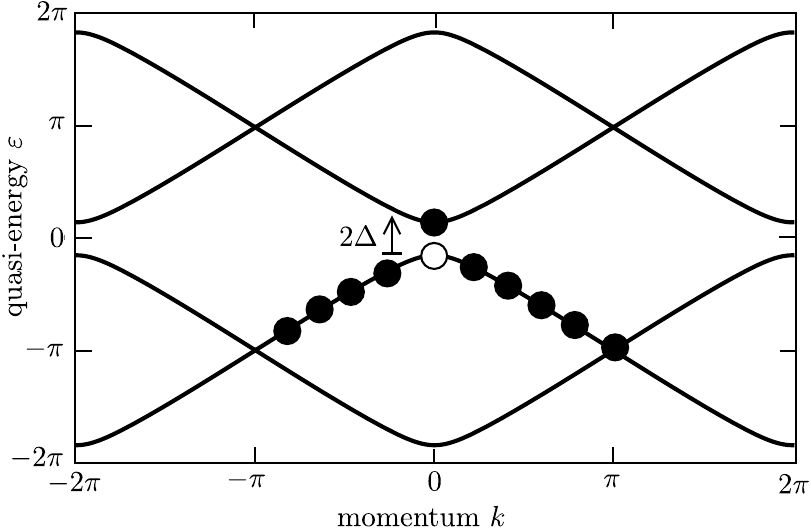}}
\caption{Dispersion relation \eqref{dispersion} of the tangent fermion quantum map \eqref{CNeq}, plotted for mass $\mu=0.5$ and velocity $v=1$. The mass does not open a gap at $k=\pi$, precluding the energy release process of Fig. \ref{fig_extended1}.
}
\label{fig_extended2}
\end{figure}

The eigenvalues $e^{-i\varepsilon}$ of ${\cal O}$ have dispersion relation
\begin{equation}
E\equiv 2\tan(\varepsilon/2)=\pm \sqrt{4v^2\tan^2(k/2)+\mu^2},
\label{dispersion}
\end{equation}
plotted in Fig.\ \ref{fig_extended2}. A quasi-energy gap $2\Delta$, with
\begin{equation}
\Delta=2\arctan(\mu/2)
\end{equation}
opens at $k=0$, while the quasi-energy band crossing at $k=\pi$ remains gapless, avoiding the Gupta-Short obstruction to a stable Dirac vacuum.

\subsection{Robustness of the gapless band crossing at the Brillouin zone corner}
\label{sec_robustness}

For $\mu=0$, $v=1$, the Floquet operator \eqref{Omegadef} reduces to ${\cal O}=e^{-ik\sigma_x}$. So at the level of the free, massless dynamics, the tangent fermion quantum map of Sec.\ \ref{sec_tangent} is identical to the Dirac quantum walk of Sec.\ \ref{sec_Dirac}. It is the response to a mass and other perturbations that is fundamentally different: For the Dirac quantum walk a mass term opens a gap both at the Brillouin zone center $(\varepsilon,k)=(0,0)$ and at the corner $(\varepsilon,k)=(\pi,\pi)$, while for tangent fermions the band crossing at the Brillouin zone corner is protected.

The protection rests on the fact that the Cayley energy $E$ is a constant of the motion of the quantum map \eqref{CNeq}. The two quasi-energy bands that cross at the Brillouin zone corner are mapped to $E=\pm\infty$, separated by an infinite gap that no bounded perturbation of ${\cal W}$ can cross.\footnote{On a finite $N$-site lattice this should be qualified, since $k=\pi$ is not in the spectrum and the largest Cayley energy is finite, $|E|_{\rm max}\simeq 4vN/\pi$ for $N\gg 1$. Then no perturbation of ${\cal W}$ that remains bounded in operator norm as $N\rightarrow\infty$ can gap the corner. This includes the linear potential ramp considered in Sec.\ \ref{sec_Schwinger}, with $\|V\|={\cal F}L$, in the limit $L\rightarrow\infty$, ${\cal F}\rightarrow 0$ at fixed ${\cal F}L$.}

The robustness of the band crossing may also be understood in terms of time averaging. Under ${\cal W}= {\cal W}_0+\delta {\cal W}$ the Cayley transform \eqref{Omegadef} has the first-order variation
\begin{equation}
\delta {\cal O}=-i\widetilde\Phi\delta{\cal W}\,\widetilde\Phi,\;\;
\widetilde\Phi\equiv\tfrac{1}{2}(1+{\cal O}).
\label{deltaUform}
\end{equation}
The perturbation enters sandwiched between two factors of the \textit{temporal} average $\widetilde\Phi$ --- the operator that averages a state before and after one step of the quantum map. On an eigenstate, $\widetilde\Phi|\chi\rangle=\tfrac{1}{2}(1+e^{-i\varepsilon})|\chi\rangle$, so the one-step transition amplitude between initial and final states $\chi_i,\chi_f$ is
\begin{equation}
|\langle\chi_f|\,\delta {\cal O}\,|\chi_i\rangle|=
|\cos(\varepsilon_f/2)\cos(\varepsilon_i/2)|
|\langle\chi_f|\delta{\cal W}|\chi_i\rangle |,
\label{deltaU}
\end{equation}
which is suppressed when $\varepsilon_i,\varepsilon_f\rightarrow\pi$.

These considerations do not apply to perturbations that do not enter ${\cal W}$ at all, such as the split-operator coupling of Sec.\ \ref{sec:gauge}, where a rapidly varying potential may open a gap at the zone corner. In the next section we continue with the fully protected implementation, returning to the split-operator implementation in Sec.\ \ref{sec:gauge}.

\section{Schwinger effect on a space-time lattice}
\label{sec_Schwinger}

The Schwinger effect \cite{Sch51,Gel16,Tay26} is a particle-hole pair production process driven by an electric field. An electron in the Fermi sea with $k<0$ is accelerated to $k>0$ by the field. As it crosses $k=0$ it can tunnel through the bandgap, forming a particle excitation of the Dirac vacuum and leaving behind a hole excitation of the Fermi sea. The interband tunneling that produces the particle-hole pair is referred to as Klein tunneling \cite{Kle29,All11,Kat06,Don22b}, first discussed in this context by Sauter \cite{Sau31,Sau32}.

\begin{figure}[tb]
\centerline{\includegraphics[width=0.9\linewidth]{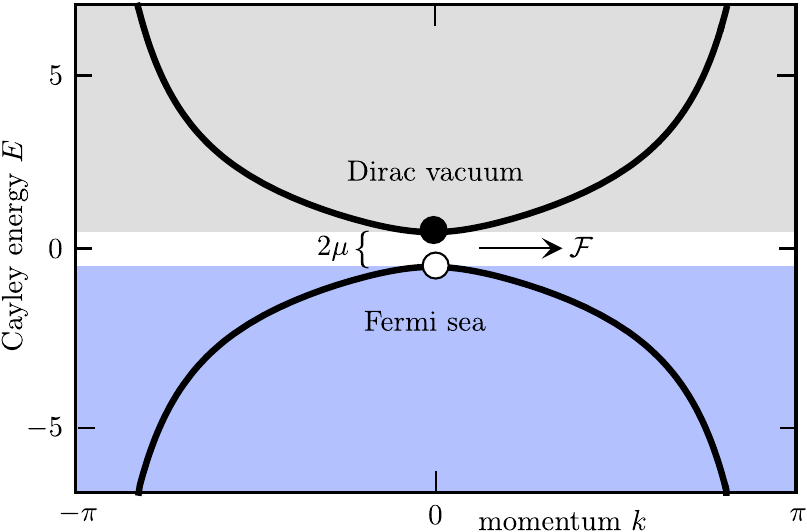}}
\caption{Dispersion relation \eqref{dispersion} plotted for the Cayley energy $E=2\tan(\varepsilon/2)$, rather than for the quasi-energy $\varepsilon$ (same parameters as in Fig.\ \ref{fig_extended2}). The regions identified with the Fermi sea and the Dirac vacuum are indicated, as well as a particle-hole pair produced by interband tunneling (Klein tunneling) in the presence of an electric field ${\cal F}$. The infinite band gap at $k=\pi$ precludes the instability mechanism of Fig.\ \ref{fig_extended1} for the spatially-averaged coupling. 
}
\label{fig_Cayley}
\end{figure}

On the space-time lattice it is helpful to visualize the Schwinger effect in terms of the Cayley energy $E$, see Fig.\ \ref{fig_Cayley}. The Cayley map of the unit circle onto the real axis allows for an unbounded Fermi sea $-\infty<E<-\mu$ and Dirac vacuum $\mu<E<\infty$. Interband tunneling then happens at $k=0$ across a $2\mu$ gap.

\subsection{Klein tunneling in an electric field}
\label{sec:kleinsetup}

We include the scalar potential $V(x)$ in the generator \eqref{Omegadef} of the tangent fermion quantum map, ${\cal W}\mapsto{\cal W}+V$. The Crank-Nicolson equation \eqref{CNeq} is then changed into
\begin{equation}
\begin{split}
&\left(\Phi^\dagger \Phi +\tfrac{1}{2}i{\cal H}\right)\Psi(t+1)=\left(\Phi^\dagger \Phi-\tfrac{1}{2}i{\cal H}\right)\Psi(t),\\
&{\cal H}=v\sigma_x\sin k+\Phi^\dagger[\mu \sigma_z +V(x)]\Phi.
\end{split}\label{HVeq}
\end{equation}
The $\Phi^\dagger V\Phi$ product spatially averages the potential over neighbouring lattice sites. This spatially-averaged coupling provides for a convenient regularization of the dynamics, but it is not gauge invariant: A constant $V=V_0$ changes the Cayley energy \eqref{dispersion} linearly,
\begin{equation}
E=V_0\pm \sqrt{4v^2\tan^2(k/2)+\mu^2},\label{Edispersion}
\end{equation}
but the change in the quasi-energy $\varepsilon=2\arctan(E/2)$ acquires nonlinear corrections. One cannot therefore remove a constant potential offset by a gauge transformation. For the calculation that follows this has no effect if we keep all energies well below the $1/\tau$ quasi-energy bandwidth, but we will also consider, in the next section, a fully gauge invariant formulation of the problem. 

We take a downward potential ramp $V(x)$, decreasing linearly from $0$ to $-{\cal F}L$ in the interval $0<x<L$, producing an electric field ${\cal F}$. The field-free regions for $x<0$ and $x>L$ allow to formulate a scattering problem (the Nikishov approach \cite{Nik70} to the Schwinger effect), with incident Fermi-sea electrons from the left (right-moving, so $k<0$) that are either reflected back into the same band, or transmitted to the right in the vacuum band (now with $k>0$). The transmission probability is $T(\varepsilon)$, in the Fermi-sea energy range $-\pi<\varepsilon<-\Delta$.

A finite observation time $t_{\rm obs}$ corresponds to discrete quasi-energies $\varepsilon_n$ with spacing $2\pi/t_{\rm obs}$. An electron incident on the potential ramp at $\varepsilon_n$ in the Fermi sea produces a particle-hole pair with probability $T_n=T(\varepsilon_n)$; with probability $1-T_n$ it leaves the Fermi sea unaffected. Since the electrons are assumed to be non-interacting, the total probability of no particle-hole production, the vacuum persistence probability $P_{\rm vac}$,  factorizes \cite{Nik70},
\begin{equation}
P_{\rm vac}=\prod_{n}(1-T_n).
\end{equation}
The resulting rate $\kappa$ that governs the exponential vacuum decay $P_{\rm vac}\propto e^{-\kappa t_{\rm obs}}$ follows in the large-$t_{\rm obs}$ limit,
\begin{align}
\kappa={}&-\lim_{t_{\rm obs}\rightarrow\infty}\frac{1}{t_{\rm obs}}\ln P_{\rm vac}\nonumber\\
={}&-\int_{-\pi}^{-\Delta}\frac{d\varepsilon}{2\pi}\ln[1-T(\varepsilon)].
\end{align}

\subsection{Klein tunneling probability for tangent fermions}
\label{sec:klein}

Substituting $\Phi\Psi(t)=e^{-i\varepsilon t}\chi$, the evolution equation \eqref{HVeq} can be rewritten identically as a Hermitian eigenvalue equation in the Cayley energy,
\begin{equation}
\;\Bigl[2v\sigma_x\tan(k/2)+\mu\sigma_z+V(x)\Bigr]\chi=E\chi.
\label{tangentdirac}
\end{equation}
We recover the dispersion relation \eqref{Edispersion} for a constant $V=V_0$.

In the large-$L$ limit we may extend the potential ramp $V(x)=-{\cal F}x$ to the whole $x$-axis and transform to momentum representation, $x=i\partial/\partial k$, when Eq.\ \eqref{tangentdirac} becomes a first order differential equation,
\begin{equation}
\begin{split}
&i{\cal F}\frac{d\chi}{dk}=H_{\text{LZ}}(k)\chi,\;\;|k|<\pi,\\
&H_{\text{LZ}}(k)=2v\sigma_x\tan(k/2)+\mu\sigma_z-E.
\end{split}
\label{LZproblem}
\end{equation}
This is a two-band Landau-Zener problem in which the momentum $k$ plays the role of time and ${\cal F}$ is the effective Planck constant.

The eigenvalues of the Landau-Zener Hamiltonian $H_{\text{LZ}}$ form two bands,
\begin{equation}
U_\pm(k)=-E\pm\sqrt{\mu^2+4v^2\tan^2(k/2)},\label{Ubands}
\end{equation}
with an avoided crossing at $k=0$. An electron starts out in band $U_-$ for $k<0$ and for $k>0$ either stays in that band or switches to band $U_+$, by tunneling through the avoided crossing. The interband tunneling probability $T$ is the Klein tunneling probability that we seek. It is independent of $\varepsilon$, since $\varepsilon$ only enters in Eq.\ \eqref{LZproblem} via $E$, and $E$ is a $k$-independent offset to $U_\pm$ which cannot affect the tunneling probability.\footnote{This $\varepsilon$-independence holds in the large-$L$ limit. In the next section we will consider a finite $L$, which introduces an energy cutoff.}

\begin{figure}[tb]
\centerline{\includegraphics[width=0.9\linewidth]{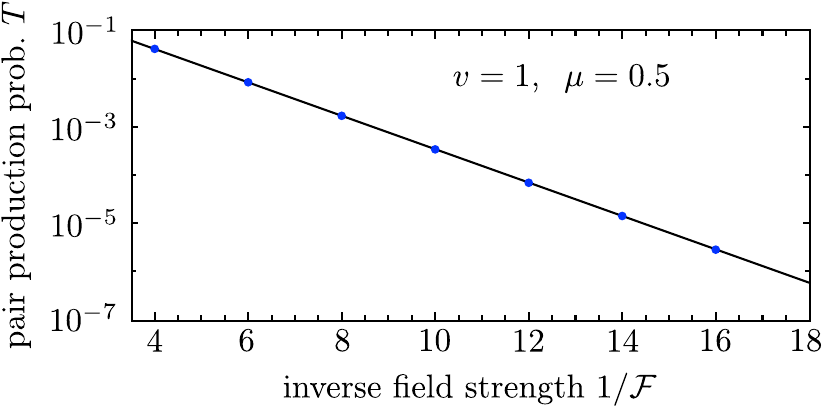}}
\caption{Pair production probability $T$ for the tangent fermion quantum map in a uniform electric field ${\cal F}$, included via the spatially-averaged coupling. The data points are computed numerically by integration of the Landau-Zener Hamiltonian \eqref{LZproblem}, the solid line is the small-${\cal F}$ result \eqref{Smain}.
}
\label{fig_numerics1}
\end{figure}

The Dykhne-Davis-Pechukas formula \cite{Dyk62,Dav76} gives the leading small-${\cal F}$ asymptotics of the transition probability,
\begin{equation}
T=\exp\Bigl[-\frac{2}{{\cal F}}\,\operatorname{Im}\!\int_{0}^{k_{c}}\bigl[U_{+}(k)-U_{-}(k)\bigr]\,dk\Bigr],\label{Dykhne}
\end{equation}
with $k_{c}$ the complex degeneracy point,
\begin{equation}
U_{+}(k_c)=U_{-}(k_c)\Rightarrow k_{c}=2i\operatorname{arctanh}(\mu/2v).\label{kc}
\end{equation}
Eq.\ \eqref{Dykhne} evaluates to
\begin{equation}
T=\exp\left[-(8\pi v/{\cal F})\left(1-\sqrt{1-(\mu/2v)^2}\right)\right],\;\;|\mu|<2 v.
\label{Smain}
\end{equation}
For $|\mu|>2 v$ an oscillatory term appears. This is a lattice artefact that we need not consider further.

Eq.\ \eqref{Smain} is an asymptotic result for ${\cal F}\ll\mu^2/v$, not an exact transition probability. A numerical check is given in Fig.\ \ref{fig_numerics1}, showing that it is reliable already for ${\cal F}\lesssim\mu^2/v$.

\subsection{Vacuum decay rate}
\label{sec:rate}

For the calculation of the Klein tunneling probability we could ignore the finite length $L$ of the voltage ramp. It enters in the vacuum decay rate by setting the energy interval where interband tunneling is allowed.

The interband transition can happen if $E-V(0)<-\mu$ (so that the electron starts out in the Fermi sea) and $E-V(L)>+\mu$ (so that it ends up in the Dirac vacuum). Thus the interval $(-\mu,\mu)$ should lie fully in the interval  $(E,E+{\cal F}L)$. This restricts $E$ to the interval $(\mu-{\cal F}L,-\mu)$, provided ${\cal F}L>2\mu$. Hence
\begin{align}
\kappa={}&-\frac{1}{2\pi}\ln(1-T)\int_{\mu-{\cal F}L}^{-\mu}\frac{d\varepsilon}{dE}\,dE\label{kapparesult}\\
={}&-\frac{1}{\pi}\ln(1-T)\bigl[\arctan[({\cal F}L-\mu)/2]-\arctan(\mu/2)\bigr].\nonumber
\end{align}

This may be simplified in the weak-field regime ${\cal F}L\ll 1$, which in terms of lattice constants means ${\cal F}L\ll 1/\tau$: the potential step should be much less than the bandwidth of the time discretization. Then the arctangents linearize,
\begin{equation}
\kappa=-\frac{{\cal F}L_{\rm eff}}{2\pi}\ln(1-T),\;\;L_{\rm eff}=L-2\mu/{\cal F}.\label{kappalinear}
\end{equation}
The difference between $L$ and $L_{\rm eff}$ can be ignored for large $L$.

Substituting Eq.\ \eqref{Smain} for the tunnel probability, and restoring lattice constants, we arrive at
\begin{widetext}
\begin{align}
\kappa={}&-\frac{{\cal F}L_{\rm eff}}{2\pi}\ln\left(1-\exp\left\{-\frac{8\pi  v}{{a^2\cal F}}\left[1-\sqrt{1-\left(\frac{a\mu}{2 v}\right)^2}\,\right]\right\}\right)\nonumber\\
={}&-\frac{{\cal F}L_{\rm eff}}{2\pi}\ln\left(1-\exp\left\{-\frac{\pi\mu^2}{ v{\cal F}}\left[1+\tfrac{1}{16}(a\mu / v)^2+{\cal O}(\mu a/ v)^4\right]\right\}\right),\label{kappafinal}
\end{align}
\end{widetext}
an expansion in the ratio of lattice constant $a$ over Compton wave length $v/\mu$. In the limit $a\rightarrow 0$ the continuum result of the 1+1-dimensional Schwinger effect is recovered \cite{Sch51,Nik70,Heb13}.

Eq.\ \eqref{kappafinal} is the rate of a single cone, with no factor of two from fermion doubling. The band crossing at the Brillouin zone corner $(\varepsilon,k)=(\pi,\pi)$ in Fig.\ \ref{fig_extended2} does not contribute to the pair production because it has an infinitely large Cayley energy band splitting,
\begin{align}
U_+(k)-U_-(k)={}&2\sqrt{\mu^2+4v^2\tan^2(k/2)}\nonumber\\
&\rightarrow\infty \;\;\text{when}\;\;k\rightarrow\pi.\label{Cayleybandsplitting}
\end{align}
Only the physical Schwinger channel at the $(\varepsilon,k)=(0,0)$ avoided crossing remains.

\section{Split-operator potential coupling}
\label{sec:gauge}

\subsection{Restoring gauge invariance}

The quantum map \eqref{HVeq} inserts the potential $V(x)$ as a \textit{spatially-averaged coupling} $\Phi^\dagger V\Phi$ (averaged over nearest neighbour sites). The potential then shifts the Cayley energy $E$ linearly, but the quasi-energy $\varepsilon$ non-linearly, leading to a voltage-offset dependence that cannot be removed by a gauge transformation. Gauge invariance is only recovered in the continuum limit, when the spurious voltage-offset dependence vanishes.

An alternative coupling that is exactly gauge invariant at any discretization is the symmetric \textit{split-operator coupling} \cite{Don22},\footnote{The \textit{symmetric} insertion of $e^{-iV/2}$, rather than a single $e^{-iV}$ operator, serves three purposes: 1) The discretization error is order $\tau^3$ rather than order $\tau^2$; 2) the Floquet operator satisfies time-reversal symmetry for $\mu=0$; 3) the scalar potential can equivalently be accounted for by time dependent Peierls phases in the hopping matrix elements (temporal gauge).}
\begin{align}
&\left(\Phi^\dagger \Phi +\tfrac{1}{2}i{\cal H}_0\right)e^{iV(x)/2}\Psi(t+1)\nonumber\\
&\qquad{}=\left(\Phi^\dagger \Phi-\tfrac{1}{2}i{\cal H}_0\right)e^{-iV(x)/2}\Psi(t),\label{CNeq2}
\end{align}
where ${\cal H}_0$ does not contain the potential,
\begin{equation}
{\cal H}_0=v\sigma_x\sin k+\Phi^\dagger\mu\sigma_z\Phi.\label{Hdef2}
\end{equation}
The corresponding Floquet operator
\begin{equation}
\tilde{\cal U}=e^{-iV/2}{\cal U}_0e^{-iV/2},\;\;{\cal U}_0=\frac{\Phi^\dagger \Phi-\tfrac{1}{2}i{\cal H}_0}{\Phi^\dagger \Phi +\tfrac{1}{2}i{\cal H}_0},
\label{Udef2}
\end{equation}
is unitary, $\tilde{\cal U}^\dagger\tilde{\cal U}=\openone$, because ${\cal H}_0$ and $\Phi^\dagger\Phi=\cos^2(k/2)$ commute (both are functions of $k$ alone).

For a linear potential, $V(x)=-{\cal F}x$, the operator $e^{-iV/2}=e^{i{\cal F}x/2}$ shifts momentum by ${\cal F}/2$, hence
\begin{equation}
\tilde{\cal U}|k\rangle= {\cal U}_0(k+{\cal F}/2)|k+{\cal F}\rangle.\label{sweep}
\end{equation}
After $n$ time steps, a momentum $k_0$ has advanced to $k_0+n{\cal F}$. This produces a spectral flow of the Fermi sea, perturbed by inelastic particle-hole pair production at $k=0$ and elastic particle-hole scattering at $k=\pi$. We consider these two processes separately.

\subsection{Pair production at the Brillouin zone center}

At $k=0$ there is an avoided crossing, the particle-hole pair requires tunneling across the $2\Delta$ quasi-energy gap of the bands
\begin{equation}
\tilde{U}_\pm(k)=V_0\pm 2\arctan\sqrt{\tfrac{1}{4}\mu^2+v^2\tan^2(k/2)}.\label{Utildebands}
\end{equation}
This is again a Landau-Zener problem, the difference with the case of spatially-averaged coupling of the previous section is that we are now tunneling through the quasi-energy gap $2\Delta=4\arctan(\mu/2)$ instead of through the Cayley-energy gap $2\mu$.

The general small-${\cal F}$ expression \eqref{Dykhne} for the transition probability \cite{Dyk62,Dav76}, now evaluated for the bands \eqref{Utildebands}, is
\begin{equation}
T_0=\exp\left(-\frac{8\mu}{v{\cal F}}\int_0^1\frac{\arctan\left(\tfrac{1}{2}\mu\sqrt{1-x^2}\right)}{1-(\mu x/2v)^2}\,dx\right),
 \label{eq:S0fixed}
\end{equation}
again for $|\mu|<2v$.
Expansion for small mass gives
\begin{equation}
T_0=\exp\biggl(-\frac{\pi\mu^2}{ v{\cal F}}\biggl[1+\tfrac{1}{16}\bigl[(a\mu/v)^2-(\tau\mu)^2\bigr]\biggr]\biggr),
 \label{eq:S0expansion}
\end{equation}
to order $\mu^2$. The space and time lattices push the exponent in opposite directions and cancel to leading order on the light-cone $a=v\tau$, leaving an order $\mu^4$ correction,
\begin{equation}
 T_0=\exp\left(-\frac{\pi\mu^2}{v{\cal F}}
 \left[1+\tfrac{5}{384}(\tau\mu)^4+{\cal O}(\mu^6)\right]\right). \label{eq:S0v1}
\end{equation}
Comparison with the result \eqref{kappafinal} for the spatially-averaged coupling shows that the split-operator coupling gives the same continuum limit, as it should, with a different discretization correction.

\subsection{Elastic scattering at the Brillouin zone corner}
\label{sec_elastic}

The quasi-energy band gap is zero at $k=\pi$, see Fig.\ \ref{fig_extended2}, so we can calculate the elastic scattering probability $T_\pi$ in perturbation theory.

We expand the one-step unitary \eqref{Udef2} near the corner, $k=\pi+p$,
\begin{equation}
\begin{split}
&{\cal U}_0(\pi+p)=-e^{-iH_\pi(p)},\\
&H_\pi(p)=\frac{p}{v}\,\sigma_{x}
-\frac{\mu p^{2}}{4v^{2}}\,\sigma_{z}+{\cal O}(p^{3}).
\end{split}
\label{Omegacorner}
\end{equation}
The linear term produces the band crossing, while the mass appears only at order $p^2$ and cannot open a gap. With $p={\cal F} t$, the effective time-dependent Hamiltonian is
\begin{equation}
 H_\pi(t)=\frac{{\cal F} t}{v}\sigma_x
 -\frac{\mu{\cal F}^2t^2}{4v^2}\sigma_z.
 \label{eq:Hpi}
\end{equation}

The $\sigma_x$ eigenstates are the $t$-independent eigenstates for $\mu=0$, representing the bands that cross at $k=\pi$. The $\sigma_z$ term couples them, and first-order perturbation theory gives the scattering amplitude
\begin{equation}
\begin{split}
& s_{\pi}=
 i\,\frac{\mu{\cal F}^2}{4v^2}
 \int_{-\infty}^{\infty}
 t^2e^{i{\cal F} t^2/v}\,dt,\\
&
\Rightarrow T_{\pi}=|s_{\pi}|^2
 =\frac{\pi(a\tau)^2\mu^2{\cal F}}{64v}+{\cal O}({\cal F}^2).
 \end{split}
 \label{eq:Pflip}
\end{equation}
With the complementary probability $1-T_\pi$ the state follows the band through the crossing --- unimpeded spectral flow of the Fermi sea, creating no excitations.\footnote{The periodicity of the spectral flow is \textit{twice} the Bloch oscillation period of $T_{\rm B}=2\pi/a{\cal F}$, because a band in Fig.\ \ref{fig_extended2} only closes on itself after two Brillouin zone traversals. In contrast, spectral flow of the Fermi sea for the Dirac quantum walk has period $T_{\rm B}$ --- the bands in Fig.\ \ref{fig_extended1} close on themselves after one zone traversal.}

The scaling $T_\pi\propto (a\tau)^2$ identifies this scattering process as a lattice artefact, which however does not compromise the vacuum stability: The electric field does work on the particle-hole pair to pull them apart, no energy is released to the field.

\begin{figure}[tb]
\centerline{\includegraphics[width=0.9\linewidth]{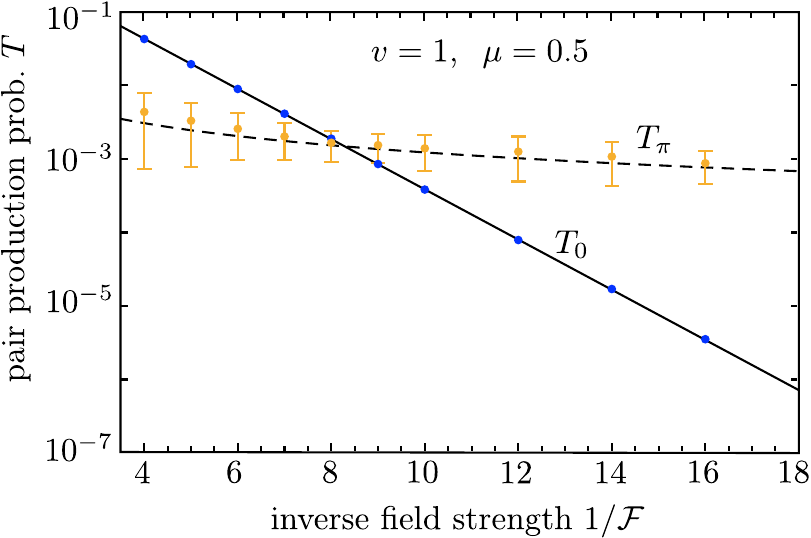}}
\caption{Pair production probability $T_0$ at zone center and $T_\pi$ at zone corner, for the tangent fermion quantum map in a uniform electric field ${\cal F}$, included via the split-operator coupling. The data points are computed numerically from the spectral flow equation \eqref{sweep}, the solid and dashed lines are the small-${\cal F}$ results \eqref{eq:S0fixed} and \eqref{eq:Pflip}, respectively. The error bars for $T_\pi$ give the standard deviation upon averaging over momentum sweeps in the interval $\pi-\delta k<k<\pi+\delta k$ with $\delta k\in(0.8,2.6)$. There is no significant sweep interval dependence for $T_0$.
}
\label{fig_numerics2}
\end{figure}

The pair production pathways from zone center and zone corner may interfere over repeated zone traversals. For $T_0,T_\pi\ll1$ and in the presence of dephasing --- or for a single traversal --- the contributions add incoherently to the pair production rate,
\begin{equation}
\kappa=\frac{{\cal F} L_{\rm eff}}{2\pi}\,(T_0+T_\pi).
\end{equation}
The ratio $T_\pi/T_0$ of the corner and center contributions vanishes $\propto(a\tau)^2$ in the continuum limit at fixed field. The field dependence of the two contributions is compared in Fig.\ \ref{fig_numerics2}. The elastic corner channel dominates at weak fields.

\subsection{Non-uniform field}
\label{sec:nonuniform}

We can relax the assumption of a uniform electric field, and consider the effect of spatial potential fluctuations. Included as a split operator in the Floquet operator \eqref{Udef2}, the potential fluctuations may open a gap $\Delta_\pi$ at the Brillouin zone corner $(\varepsilon,k)=(\pi,\pi)$. The protection of the band crossing discussed in Sec.\ \ref{sec_robustness} only applies to potentials included via the spatially-averaged coupling, not to the split-operator coupling.

Consider the potential $\delta V(x)=V_0\cos qx$, varying on the length scale of $1/q$, of vanishing first moment and second moment $\langle\delta V^2\rangle=V_0^2/2$. In App.\ \ref{app_gap} we compute the quasi-energy gap
\begin{equation}
\Delta_\pi=\frac{\mu V_0^2}{4\lambda}|J_1(2\lambda)|,\;\;\lambda=vV_0/q,
\label{besselgap2}
\end{equation}
for $\mu,V_0,q\ll 1$, as a function of the ratio $\lambda=vV_0/q$. The Bessel function interpolates between ``rough potentials'' ($\lambda\ll 1$) and ``smooth'' potentials ($\lambda\gg 1$),
\begin{equation}
\Delta_\pi=\begin{cases}
\tfrac{1}{4}\mu V_0^2,
&\lambda\ll 1,\\
\tfrac{1}{4}\mu V_0^2(\pi\lambda^3)^{-1/2}
\bigl|\cos(2\lambda+\tfrac{1}{4}\pi)\bigr|, &\lambda\gg 1.
\end{cases}\label{gapsemi2}
\end{equation}

Inserting the lattice constants, we find that if the potential $\delta V$ varies sufficiently rapidly, wave number $q\gg (\tau/a)^2 vV_0$, a gap opens of magnitude
\begin{equation}
\Delta_\pi=\tfrac{1}{2} \mu\tau^2\langle \delta V^2\rangle.\label{cornergap}
\end{equation}
The gap closes $\propto q^{3/2}$ for smooth potentials, $q\ll (\tau/a)^2 vV_0$. That is the regime where the elastic scattering calculation of Sec.\ \ref{sec_elastic} applies ($q\rightarrow 0$ at fixed $V_0$).

Once the corner is gapped the Gupta-Short instability of the Dirac quantum walk becomes operative \cite{Gup25}: A particle-hole pair releases $\Delta_\pi$ to the environment. What protects the tangent fermion vacuum is then no longer the exact decoupling of Sec.\ \ref{sec_robustness}, but the smallness of $\Delta_\pi$: The energy that can be released by a potential of amplitude $V_0$ is smaller than in the Dirac quantum walk by a factor of order $(V_0\tau)^2$, vanishing in the continuum limit.

\section{Conclusion}
\label{sec_conclude}

Gupta and Short \cite{Gup25} posed a dilemma for Dirac fermions in discrete time: \textit{either} fill the negative quasi-energy band and the vacuum can decay at the Brillouin zone corner, where pair creation releases energy $2\mu$; \textit{or} remove the band crossing at the zone corner to forbid the decay and the model doubles, with a spurious second cone at $\varepsilon=0$. 

We have shown how to resolve the dilemma by replacing the local unitary update of the quantum walk by a local, unitary, but \textit{implicit} update of the Crank-Nicolson form \cite{Cra47}. The tangent-fermion quantum map \cite{Bee23} makes the zone corner simultaneously \textit{ungappable} and \textit{unreachable}: The band crossing at the zone corner is protected from any perturbation coupled in a spatially averaged way through the Hermitian generator ${\cal W}$ of the quantum map, while in terms of the conserved Cayley energy $E=(2/\tau)\tan(\varepsilon\tau/2)$ the same crossing lies at infinity, so that pair creation there costs unbounded energy. 

The quantitative test is the Schwinger effect \cite{Sch51}: The vacuum decay rate in a uniform electric field reproduces the continuum result \cite{Nik70} for a single Dirac cone, with discretization corrections quadratic in the lattice constants, and with no doubling of the rate.

The two couplings of the potential that we compared (spatially-averaged versus split-operator) agree in the continuum limit, and they agree on the Schwinger rate in a uniform field. They differ in their discretization corrections and, more consequentially, in their stability properties. 

For the spatially-averaged coupling the protection of the zone corner is complete: No bounded perturbation of ${\cal W}$ can gap the corner or excite it, because it lies at infinite Cayley energy. The price is that gauge invariance is recovered only for $\tau\rightarrow0$. 

The split-operator coupling is exactly gauge invariant at any $\tau$, and pays for that with a corner that is no longer completely decoupled: A spatially fluctuating potential $\delta V$ opens a corner gap which allows for energy-releasing inelastic transitions -- but at a fraction $\tfrac{1}{4}\tau^2\langle\delta V^2\rangle$ of the $2\mu$ energy released in the Dirac quantum walk, and thus negligibly small for any potential well below the bandwidth $1/\tau$.

Our results show that a zone-corner crossing need not be removed in order to stabilize the Dirac vacuum. The Cayley transformation allows the crossing to remain in the quasi-energy spectrum, while placing it at infinite generator energy, beyond the reach of bounded perturbations. Exact lattice gauge invariance weakens this protection only through ultraviolet contributions that vanish in the continuum limit. Tangent fermions thereby convert the Gupta-Short dilemma from an unavoidable vacuum instability into a controllable choice of discretization, allowing single-cone Dirac physics and vacuum stability to coexist.

\acknowledgments

We used several AI models (Opus, Fable, GPT Sol) as an interactive tool to explore this topic, and for help with coding. Computer codes and data sets are available at a \href{https://doi.org/10.5281/zenodo.22209729}{Zenodo repository}.\\
This work was supported by the Netherlands Organisation for Scientific Research (NWO/OCW), as part of Quantum Limits (project number {\sc summit}.1.1016). Access to OpenAI models was provided through the ChatGPT for Academic Researchers program.

\appendix

\section{Corner gap from a non-uniform field in the split-operator coupling}
\label{app_gap}

We compute the quasi-energy gap $\Delta_\pi$ that a non-uniform field opens at the Brillouin zone corner $(\varepsilon,k)=(\pi,\pi)$, when the potential $\delta V(x)$ is coupled as a split operator, Eq.\ \eqref{Udef2}. (No gap is opened for the averaged-operator coupling, see Sec.\ \ref{sec_robustness}.)

The answer depends both on the amplitude $V_0$ of the potential and on the length scale $1/q$ over which it varies. The characteristic parameter for the smoothness of the potential\footnote{Our classification in ``rough versus smooth'' potentials focuses on the $q$ dependence of the parameter $\lambda$. Alternatively, one can focus on the $V_0$ dependence, and then distinguish ``weak versus strong'' potentials.} is $\lambda=vV_0/q$ or $\lambda=(\tau/a)^2 vV_0/q$ if we restore the lattice constants. In the next subsection we consider ``rough potentials'', $\lambda\ll 1$. Then we compute the crossover to ``smooth potentials'', $\lambda\gg 1$.

\subsection{Rough potentials}

For $\lambda\ll 1\Rightarrow V_0\ll q/v$ we may compute the corner gap by perturbation theory in the amplitude $V_0$ of the potential. 

The Floquet operator $\tilde{\cal U}=e^{-i\delta V/2}{\cal U}_0e^{-i\delta V/2}$ is related by a unitary transformation to
\begin{equation}
{\cal M}={\cal U}_0e^{-i\delta V},
\end{equation}
so the two have the same spectrum and we may work with ${\cal M}$. Let ${\cal P}$ project onto the two-dimensional spin space at $k=\pi$ and let ${\cal Q}=\openone-{\cal P}$. Since ${\cal U}_0(\pi)=-\openone$, the unperturbed eigenvalue is $-1$, twofold degenerate, and the corner gap is the splitting of that doublet.

We expand $\delta V=\sum_q v_qe^{iqx}$, with $v_{-q}=v_q^\ast$, $v_0=0$ (vanishing spatial average), and carry out the degenerate perturbation theory to second order via L\"{o}wdin partitioning,
\begin{equation}
\Lambda={\cal P}{\cal M}{\cal P}+{\cal P}{\cal M}{\cal Q}\,(u-{\cal Q}{\cal M}{\cal Q})^{-1}\,{\cal Q}{\cal M}{\cal P}.
\end{equation}
To second order we may replace the eigenvalue $u$ by its unperturbed value $u_0=-1$.

In the first term ${\cal P}{\cal U}_0{\cal P}=-\openone$ and ${\cal P}\delta V{\cal P}=v_0=0$, so the expansion $e^{-i\delta V}=\openone-i\delta V-\tfrac{1}{2}\delta V^2+\ldots$ contributes $-1+\tfrac{1}{2}\langle\delta V^2\rangle$, with $\langle\delta V^2\rangle=\sum_q|v_q|^2$ the spatial mean square of the potential. 

In the second term we may replace ${\cal Q}{\cal M}{\cal Q}$ by ${\cal Q}{\cal U}_0{\cal Q}$, to the order considered, and use the identity
\begin{equation}
(-\openone-{\cal U}_0)^{-1}{\cal U}_0=-\tfrac{1}{2}\openone+\tfrac{1}{4}i{\cal W}.
\label{keyidentity}
\end{equation}
The intermediate states lie at $k=\pi+q$, where
\begin{equation}
{\cal W}(\pi+q)=-2v\operatorname{cotan}(q/2)\,\sigma_x+\mu\sigma_z .
\end{equation}
Collecting the terms,
\begin{align}
\Lambda={}&-\openone+\tfrac{1}{2}\langle\delta V^2\rangle\openone
+\sum_q|v_q|^2\bigl[-\tfrac{1}{2}\openone+\tfrac{1}{4}i{\cal W}(\pi+q)\bigr]\nonumber\\
={}&-\openone+\tfrac{1}{4}i\,\mu\langle\delta V^2\rangle\,\sigma_z .
\end{align}
The cotangent does not contribute because it is odd in $q$.

The two states thus move from quasi-energy $\varepsilon=\pi$ to $\varepsilon=\pm(\pi-\tfrac{1}{2}\Delta_\pi)$, separated across the zone boundary by
\begin{equation}
\Delta_\pi=\tfrac{1}{2}\mu\langle\delta V^2\rangle.\label{Deltapirough}
\end{equation}
For rough potentials the gap is independent of the potential profile --- the only property of $\delta V$ that enters is its mean square.

As a check, we have computed the gap exactly for the staggered potential $\delta V=V_0(-1)^x$, alternating sign on adjacent sites, with $\langle\delta V^2\rangle=V_0^2$. We find
\begin{align}
\Delta_\pi={}&2 \arcsin\left(\frac{\mu }{\sqrt{\mu ^2+4}}\right)-2\arcsin\left(\frac{\mu  \cos V_0 }{\sqrt{\mu ^2+4}}\right)\nonumber\\
={}&\tfrac{1}{2}\mu V_0^2+{\cal O}(V_0^4),
\end{align}
in agreement with Eq.\ \eqref{Deltapirough}.

\subsection{Smooth potentials}

The gap \eqref{Deltapirough} holds in the rough-potential regime $\lambda\ll 1$. We now relax that assumption and compute the entire crossover from rough to smooth potentials. For that purpose we may assume $q\ll 1$ ($aq\ll 1$), since the result \eqref{Deltapirough} already applies for potentials that vary rapidly on the scale of the lattice constant. 

We compute $\Delta_\pi$ to first order in $\mu$, for $q,V_0\ll 1$ at arbitrary ratio $\lambda=vV_0/q$. Our assumption $q\ll 1$ allows us to represent $p=-i\partial/\partial x$ by a differential operator.

We start by expanding the Floquet operator $\tilde{U}=e^{-i\delta V/2}{\cal U}_0e^{-i\delta V/2}$ about the corner $k=\pi+p$,
\begin{equation}
\begin{split}
&\tilde{U}=-e^{-i\delta V/2}e^{-iH_\pi(p)}e^{-i\delta V/2},\\
&H_\pi(p)=\frac{p}{v}\,\sigma_{x}
-\frac{\mu p^{2}}{4v^{2}}\,\sigma_{z}+{\cal O}(p^{3}).
\end{split}
\end{equation}
We combine the exponentials into an effective Hamiltonian, $\tilde{U}=-e^{-iH_{\rm eff}}$,
\begin{align}
H_{\rm eff}={}&H_\pi+\delta V+\\
&+\tfrac{1}{24}[\delta V,[\delta V,H_\pi]]
-\tfrac{1}{12}[H_\pi,[H_\pi,\delta V]]+\cdots . \nonumber
\end{align}
The nested commutators, of order $qV_0^2$ and $q^2 V_0$, are of third order in the small parameters $q,V_0$ and we neglect them.

For convenience we rename the Pauli matrices so that the chirality is diagonal: $\sigma_x\mapsto\tau_z$, $\sigma_z\mapsto\tau_x$. The quasi-energy $\varepsilon=\pi+\delta\varepsilon$ then follows from the eigenvalue problem
\begin{equation}
\begin{split}
&H_{\rm eff}\psi=\bigl[(p/v)\tau_z+m(p)\tau_x+\delta V(x)\bigr]\psi=\delta\varepsilon\psi,\\
&p=-i\partial/\partial x,\;\;m(p)=-\tfrac{1}{4}\mu(p/v)^2.
\end{split}
\label{corner_continuum}
\end{equation}

For $\mu=0$ Eq.\ \eqref{corner_continuum} at eigenvalue $\delta\varepsilon=0$ has two degenerate eigenstates ${{\psi_-}\choose{0}}$, ${{0}\choose{\psi_+}}$ with 
\begin{equation}
\psi_{\pm}=\ell^{-1/2}e^{\pm i\varphi(x)},\;\;
\varphi(x)=v\int_0^x \delta V(x')\,dx' .
\label{RLstates}
\end{equation}
The eigenstates are normalized over the period $\ell=2\pi/q$ of the potential. Continuity is assured by $\int_0^\ell \delta V(x)\,dx=0$.

To find the splitting to first order in $\mu$, we apply degenerate perturbation theory in the subspace of the zero-energy eigenvalues, resulting in
\begin{equation}
\Delta_\pi=2\bigl|\langle\psi_{-}|m(p)|\psi_{+}\rangle\bigr|
=\frac{2}{\ell}\biggl|\int_0^\ell e^{i\varphi}\,m(p)\,e^{i\varphi}\,dx\biggr|,
\label{Deltadef}
\end{equation}
up to corrections of order $\mu^2$.

Using $\partial_xe^{i\varphi}=iv\delta V e^{i\varphi}$ twice in Eq.\ \eqref{Deltadef}, we have
\begin{align}
\int_0^\ell e^{i\varphi}\,m(p)\,e^{i\varphi}\,dx={}&-\tfrac{1}{4}\mu\int_0^\ell\bigl[\delta V^2-iv^{-1}\delta V'\bigr]e^{2i\varphi}\,dx\nonumber\\
={}&\tfrac{1}{4}\mu\int_0^\ell \delta V^2e^{2i\varphi}\,dx,
\end{align}
the second equality because $-iv^{-1}\delta V'e^{2i\varphi}=-i\,\partial_x\bigl(v^{-1}\delta Ve^{2i\varphi}\bigr)-2\delta V^2e^{2i\varphi}$, and the total derivative integrates to zero over a period. We thus arrive at the gap equation
\begin{equation}
\Delta_\pi=\frac{\mu}{2\ell}\biggl|\int_0^\ell \delta V^2e^{2i\varphi}\,dx\biggr|.\label{Deltamaster}
\end{equation}

\begin{figure}[tb]
\centerline{\includegraphics[width=0.95\linewidth]{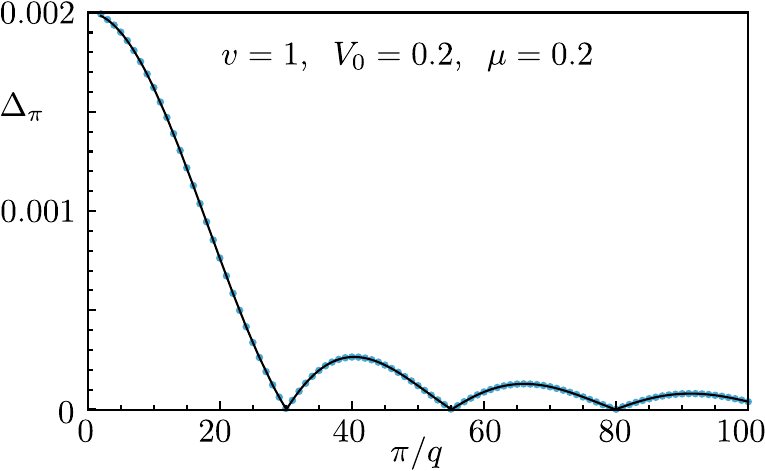}}
\caption{Corner gap of the tangent fermion quantum map, produced by a potential $\delta V=V_0\cos qx$ inserted as a split operator, $\tilde{\cal U}=e^{-i\delta V/2}{\cal U}_0e^{-i\delta V/2}$. Data points are computed numerically by diagonalization of $\tilde{\cal U}$, the curve is the small $q,\mu,V_0$ asymptotic \eqref{gapsemi}. The gap decays $\propto q^{3/2}$ as the potential becomes smoother and smoother.
}
\label{fig_cornergap}
\end{figure}

This expression holds for any periodic potential $\delta V(x)$ with zero mean. For $\delta V=V_0\cos qx$ the integral may be evaluated in closed form. One then has $\varphi=(vV_0/q)\sin qx=\lambda\sin qx$, and Eq.\ \eqref{Deltamaster} reduces to a Bessel function,
\begin{equation}
\Delta_\pi=\frac{\mu V_0^2}{4\lambda}|J_1(2\lambda)| .
\label{besselgap}
\end{equation}
The entire crossover for a cosine potential is thus a single Bessel factor, interpolating between the ``rough'' and ``smooth'' regimes,
\begin{equation}
\Delta_\pi=\tfrac{1}{4}\mu V_0^2\times\begin{cases}
\bigl(1-\tfrac{1}{2}\lambda^2+\tfrac{1}{12}\lambda^4+\cdots\bigr),
&\lambda\ll 1,\\
(\pi\lambda^3)^{-1/2}
\bigl|\cos(2\lambda+\tfrac{1}{4}\pi)\bigr|, &\lambda\gg 1.
\end{cases}\label{gapsemi}
\end{equation}
For a cosine potential $\langle\delta V^2\rangle=\tfrac{1}{2}V_0^2$, so the $\lambda\rightarrow 0$ limit of this crossover formula agrees with the rough limit \eqref{Deltapirough}. Fig.\ \ref{fig_cornergap} shows that the asymptotic result \eqref{besselgap} accurately describes the numerical data over the entire regime from rough to smooth potentials.

\begin{thebibliography}{99}
\bibitem{Ven12} S. E. Venegas-Andraca, \textit{Quantum walks: a comprehensive review}, Quantum Inf. Proc. \textbf{11}, 1015 (2012).
\bibitem{Fau24} B. Fauseweh, \textit{Quantum many-body simulations on digital quantum computers: State-of-the-art and future challenges}, Nature Comm. \textbf{15}, 2123 (2024).
\bibitem{Mor23} T. Mori, \textit{Floquet states in open quantum systems}, Ann. Rev. Cond. Matt. Phys. \textbf{14}, 35 (2023).
\bibitem{Laz14} A. Lazarides, A. Das, and R. Moessner, \textit{Equilibrium states of generic quantum systems subject to periodic driving}, Phys. Rev. E \textbf{90}, 012110 (2014).
\bibitem{Ale14} L. D'Alessio and M. Rigol, \textit{Long-time behavior of isolated periodically driven interacting lattice systems}, Phys. Rev. X \textbf{4}, 041048 (2014).
\bibitem{Pon15} P. Ponte, A. Chandran, Z. Papi\'{c}, and D. A. Abanin, \textit{Periodically driven ergodic and many-body localized quantum systems}, Ann. Physics \textbf{353}, 196 (2015).
\bibitem{Buk15} M. Bukov, L. D'Alessio, and A. Polkovnikov, \textit{Universal high-frequency behavior of periodically driven systems: from dynamical stabilization to Floquet engineering}, Adv. Phys. \textbf{64}, 139 (2015).
\bibitem{Hey19} M. Heyl, P. Hauke, and P. Zoller, \textit{Quantum localization bounds Trotter errors in digital quantum simulation}, Sci. Adv. \textbf{5}, eaau8342 (2019).
\bibitem{Sie19} L. M. Sieberer, T. Olsacher, A. Elben, M. Heyl, P. Hauke, F. Haake, and P. Zoller, \textit{Digital quantum simulation, Trotter errors, and quantum chaos of the kicked top}, npj Quantum Inf. \textbf{5}, 78 (2019).
\bibitem{Aba16} D. A. Abanin, W. De Roeck, and F. Huveneers, \textit{Exponentially slow heating in periodically driven many-body systems}, Phys. Rev. Lett. \textbf{115}, 256803 (2015).
\bibitem{Mor16} T. Mori, T. Kuwahara, and K. Saito, \textit{Rigorous bound on energy absorption and generic relaxation in periodically driven quantum systems}, Phys. Rev. Lett. \textbf{116}, 120401 (2016).
\bibitem{Aba17} D. A. Abanin, W. De Roeck, W. W. Ho, and F. Huveneers, \textit{Effective Hamiltonians, prethermalization, and slow energy absorption in periodically driven many-body systems}, Phys. Rev. B \textbf{95}, 014112 (2017).
\bibitem{Fey65} R. P. Feynman and A. R. Hibbs, \textit{Quantum Mechanics and Path Integrals} (McGraw-Hill, New York, 1965).
\bibitem{Gup25} C. Gupta and A. J. Short, \textit{The Dirac vacuum in discrete spacetime}, Quantum \textbf{9}, 1845 (2025).
\bibitem{Jol23} N. Jolly and G. Di Molfetta, \textit{Twisted quantum walks, generalised Dirac
equation and fermion doubling}, Eur. Phys. J. D \textbf{77}, 80 (2023).
\bibitem{Gup26} C. Gupta and A. J. Short, \textit{Fermion doubling in Dirac quantum walks}, Phys. Rev. A \textbf{114}, 012208 (2026).
\bibitem{Bes21} T. Bessho and M. Sato, \textit{Nielsen-Ninomiya theorem with bulk topology: duality in Floquet and non-Hermitian systems}, Phys. Rev. Lett. \textbf{127}, 196404 (2021).
\bibitem{Bee26} C. W. J. Beenakker, J. S\'{a}nchez F\'{e}rnan, and J. Tworzyd{\l}o, \textit{Fragile single-cone Dirac quantum walks in two dimensions}, arXiv:2607.05112.
\bibitem{Bee23} C. W. J. Beenakker, A. Don\'{i}s Vela, G. Lemut, M. J. Pacholski, and J. Tworzyd{\l}o, \textit{Tangent fermions: Dirac or Majorana fermions on a lattice without fermion doubling}, Ann.\ Physik \textbf{535}, 2300081 (2023).
\bibitem{Sta82} R. Stacey, \textit{Eliminating lattice fermion doubling}, Phys. Rev. D \textbf{26}, 468 (1982).
\bibitem{Ben83} C. M. Bender, K. A. Milton, and D. H. Sharp, \textit{Consistent formulation of fermions on a Minkowski lattice}, Phys.\ Rev.\ Lett.\ \textbf{51}, 1815 (1983).
\bibitem{Pac21} M. J. Pacholski, G. Lemut, J. Tworzyd{\l}o, and C. W. J. Beenakker,
\textit{Generalized eigenproblem without fermion doubling for Dirac fermions on a lattice}, SciPost Phys. \textbf{11}, 105 (2021).
\bibitem{Cra47} J. Crank and P. Nicolson, \textit{A practical method for numerical evaluation of solutions of partial differential equations of the heat conduction type}, Math. Proc. Cambridge Phil. Soc. \textbf{43}, 50 (1947).
\bibitem{Don22} A. Don\'{i}s Vela, M. J. Pacholski, G. Lemut, J. Tworzyd{\l}o, and C. W. J. Beenakker, \textit{Massless Dirac fermions on a space-time lattice with a topologically protected Dirac cone}, Ann.\ Physik \textbf{534}, 2200206 (2022).
\bibitem{Sch51} J. Schwinger, \textit{On gauge invariance and vacuum polarization}, Phys. Rev. \textbf{82}, 664 (1951).
\bibitem{Gel16} F. Gelis and N. Tanji, \textit{Schwinger mechanism revisited}, Prog. Part. Nucl. Phys. \textbf{87}, 1 (2016).
\bibitem{Tay26} H. Taya, \textit{Schwinger effect in QCD and nuclear physics}, arXiv:2603.07847.
\bibitem{Str06} F. W. Strauch, \textit{Relativistic quantum walks}, Phys. Rev. A \textbf{73}, 054302 (2006).
\bibitem{Sun12} T. Sunada and T. Tate, \textit{Asymptotic behavior of quantum walks on the line}, J. Funct. Anal. \textbf{262},  2608 (2012).
\bibitem{Arr14} P. Arrighi, V. Nesme, and M. Forets, \textit{The Dirac equation as a quantum walk: higher dimensions, observational convergence}, J. Phys. A \textbf{47}, 465302 (2014).
\bibitem{Kle29} O. Klein, \textit{Die Reflexion von Elektronen an einem Potentialsprung nach der relativistischen Dynamik von Dirac}, Z. Physik \textbf{53}, 157 (1929).
\bibitem{All11} P.E. Allain and J.N. Fuchs, \textit{Klein tunneling in graphene: optics with massless electrons}, Eur. Phys. J. B \textbf{83}, 301 (2011).
\bibitem{Kat06} M. I. Katsnelson, K. S. Novoselov, and A. K. Geim, \textit{Chiral tunnelling and the Klein paradox in graphene}, Nature Phys. \textbf{2}, 620 (2006).
\bibitem{Don22b} A. Don\'{i}s Vela, G. Lemut, M. J. Pacholski, J. Tworzyd{\l}o, and C. W. J. Beenakker, \textit{Reflectionless Klein tunneling of Dirac fermions: comparison of split-operator and staggered-lattice discretization of the Dirac equation}, J. Phys. Condens. Matter \textbf{34}, 364003 (2022).
\bibitem{Sau31} F. Sauter, \textit{\"{U}ber das Verhalten eines Elektrons im homogenen elektrischen Feld nach der relativistischen Theorie Diracs}, Z. Physik \textbf{69}, 742 (1931).
\bibitem{Sau32} F. Sauter, \textit{Zum Kleinschen Paradoxon}, Z. Physik \textbf{73}, 547 (1932).
\bibitem{Nik70} A. I. Nikishov, \textit{Barrier scattering in field theory: removal of Klein paradox}, Nucl. Phys. B \textbf{21}, 346 (1970).
\bibitem{Dyk62} A. M. Dykhne, \textit{Adiabatic perturbation of discrete spectrum states}, Sov. Phys. JETP \textbf{14}, 941 (1962).
\bibitem{Dav76} J. P. Davis and P. Pechukas, \textit{Nonadiabatic transitions induced by a time-dependent Hamiltonian in the semiclassical/adiabatic limit: the two-state case}, J. Chem. Phys. \textbf{64}, 3129 (1976).
\bibitem{Heb13} F. Hebenstreit, J. Berges, and D. Gelfand, \textit{Simulating fermion production in 1+1 dimensional QED}, Phys. Rev. D \textbf{87}, 105006 (2013).

\end{thebibliography}
\end{document}